**JWST observations of stellar occultations by solar system bodies and rings**

P. Santos-Sanz[1], R. G. French[2], N. Pinilla-Alonso[3], J. Stansberry[4], Z-Y. Lin[5], Z-W. Zhang[6], E. Vilenius[7,8], Th. Müller[7], J.L. Ortiz[1], F. Braga-Ribas[9], A. Bosh[10], R. Duffard[1], E. Lellouch[11], G. Tancredi[12], L. Young[13] , S. N. Milam[14] and the JWST 'occultations' focus group.

(1) Instituto de Astrofísica de Andalucía, IAA-CSIC, Glorieta de la Astronomía s/n, 18008 Granada, Spain. psantos@iaa.es
(2) Department of Astronomy, Wellesley College, Wellesley, MA 02481, USA.
(3) Department of Earth and Planetary Sciences, University of Tennessee, Knoxville, 37996 TN, USA
(4) Space Telescope Science Institute, 3700 San Martin Drive, Baltimore, MD 21218, USA
(5) Institute of Astronomy, National Central University, Jhongli District, Taoyuan City 32001, Taiwan
(6) Institute of Astronomy and Astrophysics, Academia Sinica, P.O. Box 23-141, Taipei 10617, Taiwan
(7) Max Planck Institute for extraterrestrial Physics, Garching, Germany
(8) Max Planck Institute for Solar System Research, Göttingen, Germany.
(9) Federal University of Technology - Paraná (UTFPR / DAFIS), Curitiba, Brazil
(10) Department of Earth, Atmospheric, and Planetary Sciences, MIT, Cambridge, MA, USA
(11) LESIA-Observatoire de Paris, CNRS, UPMC Univ. Paris 6, Univ. Paris-Diderot, Meudon, France
(12) Departamento de Astronomía, Facultad de Ciencias, Montevideo, Uruguay
(13) SwRI, 1050 Walnut St., Boulder, CO 80302-5150, USA
(14) NASA Goddard Space Flight Center, Greenbelt, MD, USA

**-Abstract**

In this paper we investigate the opportunities provided by the James Webb Space Telescope (JWST) for significant scientific advances in the study of solar system bodies and rings using stellar occultations. The strengths and weaknesses of the stellar occultation technique are evaluated in light of JWST's unique capabilities. We identify several possible JWST occultation events by minor bodies and rings, and evaluate their potential scientific value. These predictions depend critically on accurate *a priori* knowledge of the orbit of JWST near the Sun-Earth Lagrange-point 2 (L2). We also explore the possibility of serendipitous stellar occultations by very small minor bodies as a by-product of other JWST observing programs. Finally, to optimize the potential scientific return of stellar occultation observations, we identify several characteristics of JWST's orbit and instrumentation that should be taken into account during JWST's development.

**Keywords:** Solar System: Planets, Rings, Minor Bodies. Astronomical Techniques: Stellar occultations, photometry.

## 1.- Introduction

It has been more than 20 years since the discovery of the first Trans-Neptunian Object (TNO)[1]. In the intervening years, more than 1500 TNOs and Centaurs have been discovered, and this is still only a small fraction of the estimated total population. The study of this population using visible, near-infrared, and thermal photometry and spectroscopy has provided significant knowledge of the outer solar system (e.g. Barucci et al. 2008; Santos-Sanz et al. 2009; Brown et al. 2012; Lellouch et al. 2013). Nevertheless, we are not yet able to translate the conditions in the outer nebula during its various stages (planet formation, subsequent growth and orbital evolution, and physical and chemical changes to surfaces) into a clear picture of the chemical, dynamical, and thermal history of the outer solar system. A crucial missing piece to this puzzle is the size and albedo distribution of small bodies in the outer solar system.

Only about 140 TNOs and Centaurs have diameter and albedo determinations, based in most cases on the radiometric technique applied to Spitzer and Herschel observations (e.g. Stansberry et al. 2008; Lellouch et al. 2013). These values are typically accurate at best to 10 % in diameter and 20 % in albedo, significantly restricting our detailed understanding of the intrinsic variability in the surface properties and sizes of these objects. The occultation technique is far more powerful because under optimal circumstances it can provide sizes and shapes to an accuracy of about 0.1 % (Sicardy et al. 2011; Ortiz et al. 2012).

In the case of binary objects, the mass (determined through Kepler's law) together with the size allows the determination of the bulk density. Since radiometric diameters are accurate to about 10 %, this corresponds to a 30 % uncertainly in the bulk density. Density estimates have been made for only 16 objects in the trans-Neptunian region (*e.g.* Grundy et al. 2007; Noll et al. 2008; Tholen et al. 2008; Benecchi et al. 2010; Sicardy et al. 2011; Brown et al. 2010 and 2012; Grundy et al. 2012; Stansberry et al 2012; Mommert et al. 2012; Santos-Sanz et al. 2012; Vilenius et al. 2012 and 2014; Brown 2013; Fornasier et al. 2013; Grundy et al. 2015). In a few cases (e.g. Eris, Makemake) stellar occultations have vastly reduced the diameter (and therefore density) uncertainty. Because bulk density is a critical quantity for our understanding of the interior of these bodies, the targets with occultation-derived sizes and densities play a critical role in refining our knowledge of processes in the outer solar nebula (and in assessing the reliability of the much larger sample of radiometric sizes).

Historically, occultation observations have also resulted in significant serendipitous discoveries, such as the presence of rings around small bodies (Braga-Ribas et al. 2014; Duffard et al. 2014; Ortiz et al. 2015), or even the presence of an atmosphere (Elliot et al. 1989; Elliot & Young 1992; Sicardy et al. 2011; Ortiz et al. 2012). A key scientific objective of JWST occultation observations is the search for TNO atmospheres. Although models for TNOs suggest that most of them do not retain large amounts of volatiles ($N_2$, CO, $CO_2$, $CH_4$) (Schaller & Brown 2007), local atmospheres may well be present (Ortiz et al. 2012). The threshold surface pressure that can be detected with a stellar occultation of very high SNR is below the nbar level. As side products of the stellar occultation, the orbit of the TNO can be refined as well.

[1]Pluto was not considered to be a TNO in 1992 when (15760) 1992 QB1 was discovered by Jewitt & Luu 1993.

Historical occultation observations by Pluto (Hubbard et al. 1988; Elliot et al. 1989; Elliot & Young 1992 and references therein) illustrate the power of this observing technique to detect tenuous atmospheres and to monitor seasonal variations in atmospheric structure and surface pressure (Elliot 1979). The record of the light curve during an occultation enables us to study the vertical profile of the atmosphere, which is key to understanding the equilibrium of the different species of ices on the surface of the body. Pluto is also the target of the NASA *New Horizons* mission, which will arrive at the system on 14 July 2015. Undoubtedly, our understanding of this extraordinary icy object will be transformed by the *New Horizons* observations, but this detailed view will be restricted to a snapshot in time of a system that has proven to be highly variable. Near-infrared spectroscopy from ground-based telescopes and Hubble Space Telescope (HST) light curve monitoring of Pluto, performed over the last 30 years, have helped to clarify the picture of this dynamic system (Grundy et al. 2013; Buie et al. 2010). The timescale of this variability is on the order of only months to years, a somewhat surprising result given the low temperatures and large distance from the sun. Additional observations of the Pluto system after the *New Horizons* encounter will be essential in order to understand Pluto's seasonal cycle in detail, and JWST can play a critical role in these studies.

As important as occultations are for the knowledge of Trans-Neptunian Objects, this is a relatively new field (apart from occultations of Pluto that deserve a detailed and separate discussion). Only ~ 17 occultations by 9 TNOs (see Table 1) have been recorded so far from ground-based telescopes, highlighting the importance of identifying opportunities with other facilities, including SOFIA or space-based telescopes such as HST and JWST.

Stellar occultations are also a powerful tool to explore the outer solar system, where faintness and small angular diameters prevent us from building a complete census of the objects. In principle, it is possible to explore populations of small objects whose sky-plane density is large enough to produce serendipitous stellar occultations. From such occultations, we can detect tiny objects invisible in direct imaging (e.g. Roques et al. 2008). The method can be successful if the number density of objects is sufficient to result in a significant number of observable events. This is a powerful technique to determine not only the size of the smaller TNOs but also to constrain the size-frequency distribution (an important tracer of the original population), the mechanisms of depletion, and the collisional state of the population. Additionally, the recent exciting discovery and characterization of rings around tiny Centaurs Chariklo (Braga-Ribas et al. 2014; Duffard et al. 2014) and Chiron (Ortiz et al. 2015) suggest opportunities for finding additional ring systems in the solar system.

A third important area of research using the technique of stellar occultations is the study of rings around planets. Indeed, the ring systems of Uranus and Neptune were first discovered using stellar occultations (Elliot et al. 1977; Guinan et al. 1982; Hubbard et al. 1985). Early occultation observations of Saturn's rings (Bobrov 1963) gave tantalizing hints of the structure of the ring system, but the full complexity of the ring system at the sub-km resolution scale has only been revealed by the accumulation of hundreds of stellar and radio occultation observations by the *Voyager* and *Cassini* spacecraft, complemented by HST and earth-based occultation observations. The ring systems of the giant planets all show time-variable behavior that can be monitored using occultations from JWST.

Both predicted and serendipitous stellar occultations require high-SNR, high time resolution observations of target stars (Roques et al. 2009). JWST can provide rapid observing cadences even for faint stars, as we discuss in more detail in the next sections.

Table 1. Recent stellar occultations by TNOs.

| TNO | Date | Location | Diameter (km) | Density ($g \cdot cm^{-3}$) | $D_{equiv}$ (km)[§] |
|---|---|---|---|---|---|
| 2002 $TX_{300}$ | 9 Oct, 2009 | Hawaii, multi | 286 ± 10[(a)] | ___ | ___ |
| Varuna | 19 Feb, 2010 | Brazil, single | >1003 ± 9[(b)] | ___ | 668+154-86 |
| Eris | 6 Nov, 2010 | Chile, multi | 2326 ± 12[(c)] | 2.52 ± 0.05[(c)] | ___ |
| 2003 $AZ_{84}$ | 8 Jan, 2011 | Chile, single | > 573 ± 21[(d)] | ___ | 727+62-67 |
| Quaoar | 11 Feb, 2011 | USA, single | > 760[(e)] | ___ | 1074 ± 38 |
| Makemake | 23 Apr, 2011 | Chile, Brazil, multi | 1430 ± 9[(f)] | 1.7 ± 0.3[(f)] | ___ |
| Quaoar | 4 May, 2011 | Chile, Brazil, multi | 1110 ± 5[(g)] | 1.99 ± 0.42[(g)] | 1074 ± 38 |
| 2003 $AZ_{84}$ | 3 Feb, 2012 | Israel, India, multi | 686 ± 14[(h)] | 0.76+0.30-0.17[(h)] | 727+62-67 |
| Quaoar | 17 Feb, 2012 | France, single | ~1370[(g)] | 1.99 ± 0.42[(g)] | 1074 ± 38 |
| 2002 $KX_{14}$ | 26 Apr, 2012 | Spain, single | >415 ± 1[(i)] | ___ | 455 ± 27 |
| Quaoar | 15 Oct, 2012 | Chile, single | > 400[(g)] | ___ | 1074 ± 38 |
| Varuna* | 8 Jan, 2013 | Japan, multi[(1,2,3)] | ___ | ___ | 668+154-86 |
| Sedna* | 13 Jan, 2013 | Australia, single[(1,2)] | ___ | ___ | 906+314-258 |
| Quaoar | 8 Jul, 2013 | Venezuela, single | > 1138 ± 25[(j)] | < 1.82 ± 0.28[(j)] | 1074 ± 38 |
| 2003 $AZ_{84}$* | 2 Dec, 2013 | Australia, single[(2,3)] | ___ | ___ | 727+62-67 |
| 2003 $VS_2$* | 12 Dec, 2013 | Reunion, single[(2,3)] | ___ | ___ | 523+35-34 |
| Varuna* | 11 Feb, 2014 | Chile, multi[(2,3)] | ___ | ___ | 668+154-86 |
| 2003 $VS_2$* | 4 Mar, 2014 | Israel, single[(2,3)] | ___ | ___ | 523+35-34 |
| Orcus' satellite* | 1 Mar, 2014 | Japan, single[(2)] | ___ | ___ | 276 ± 17[(k)] |
| Ixion* | 24 Jun, 2014 | Australia, single[(3)] | ___ | ___ | 617+19-20 |
| 2003 $VS_2$* | 7 Nov, 2014 | Argentina, multi[(6)] | ___ | ___ | 523+35-34 |
| 2007 $UK_{126}$* | 15 Nov, 2014 | USA, multi[(4,5)] | ___ | ___ | ___ |
| 2003 $AZ_{84}$* | 15 Nov, 2014 | Japan, China, Thailand, multi[(6)] | ___ | ___ | 727+62-67 |

Recent stellar occultations by TNOs and observer locations, indicating whether the observations resulted in a single chord or multiple chords. Updated and adapted table from Ortiz et al. 2014. §Equivalent diameters obtained via the radiometric technique using Herschel Space Observatory thermal data within the "TNOs are Cool" project (Lellouch et al. 2013) for comparison with the stellar occultation diameters. *Events not completely analyzed or not published yet: (1) Benedetti-Rossi et al. 2014a; (2) Ortiz et al. 2014; (3) Benedetti-Rossi et al. 2014b; (4) Benedetti-Rossi et al. 2015; (5) Buie et al. 2015; (6) Ortiz, private communication 2015. **References:** (a) Elliot et al. 2010; (b) Sicardy et al. 2010; (c) Sicardy et al. 2011; (d) Braga-Ribas et al.2011; (e) Person et al. 2011; (f) Ortiz et al. 2012; (g) Braga-Ribas et al. 2013; (h) Braga Ribas et al. 2012; (i) Alvarez-Candal et al. 2014; (j) Davis et al. 2014; (k) Fornasier et al. 2013.

## 2.- Science objectives and requirements

### 2.1.-Predictable stellar occultations

Predictable stellar occultations (unlike serendipitous occultations) are those that can be anticipated, i.e. it is possible to estimate when the occultation will occur, and where it will be visible. Successful occultation predictions require: i) accurate knowledge of the target orbit and ephemeris; ii) the (approximate) size of the body; iii) an accurate catalog of the positions and apparent magnitudes of stars in the vicinity of the occultation track; and (iv) accurate knowledge of the location of potential observers (this is very relevant, in particular, for space-based observatories).

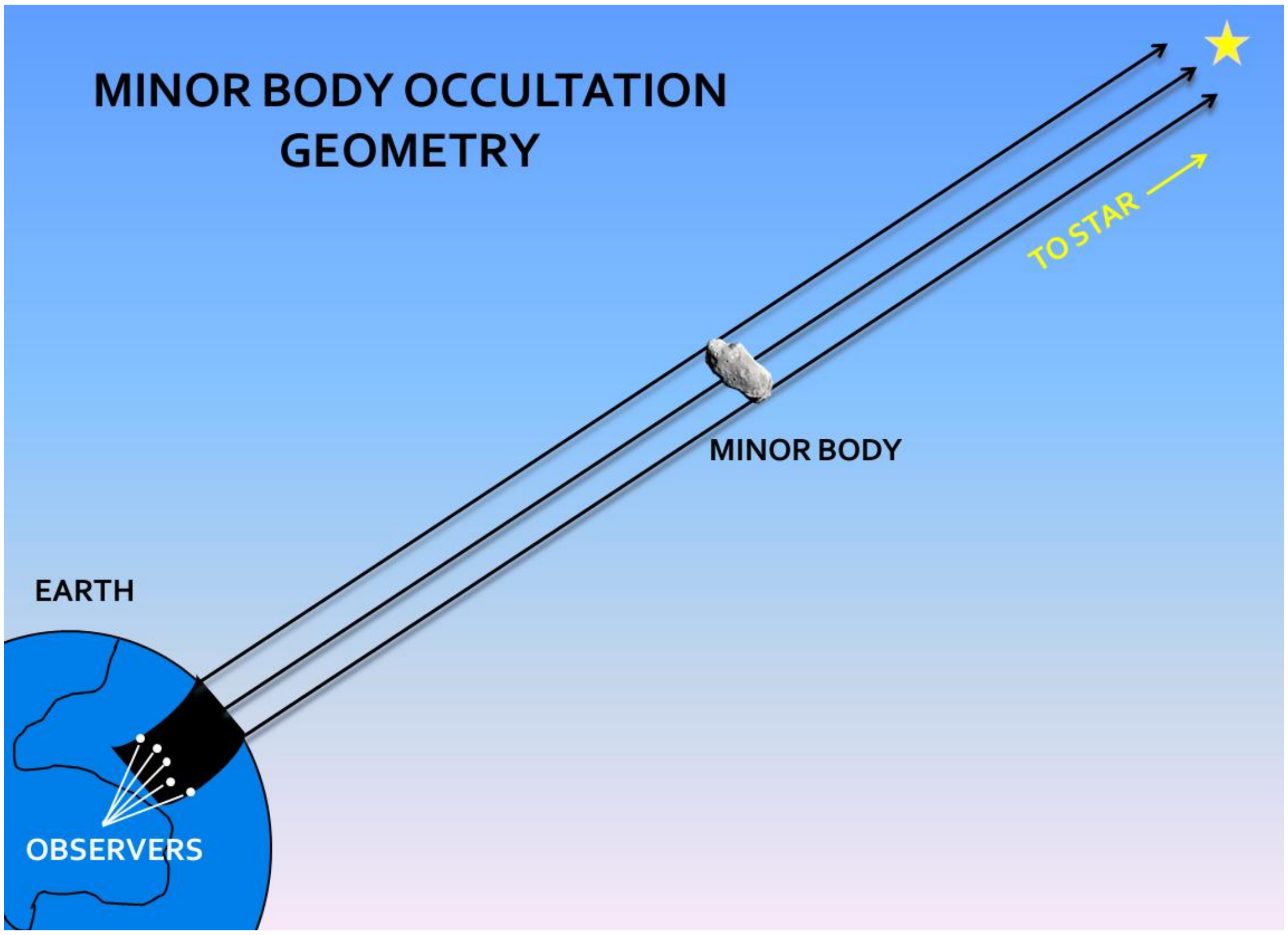

Figure 1.- Minor body occultation geometry. Scheme of a stellar occultation by a solar system minor body. The occulted star is so far away compared with the earth-body distance that it is effectively at infinity, therefore, the shadow-size projected on the earth surface matches the size of the body. If occultation is observed from different locations on the earth (white dots into the shadow) we can obtain different chords and recover the size and shape of the minor body.

For objects with very well determined orbits (whether they are major planets, satellites or numbered asteroids/TNOs), it is relatively easy to predict whether an object will occult a particular star. However, the orbits of outer solar system minor bodies, including the Centaurs (whose orbits cross the orbital distances of one or more giant planets) and the TNOs (and their sub-populations) require centuries or more to complete, so we have observed only a small arc of their orbits. As a result the orbital elements are typically quite poorly known for these objects (Virtanen et al. 2008). Apart from this, the angular size of an object is the most important factor limiting the accuracy of stellar occultation predictions. For objects with an angular size larger than that of the star they occult (typically around 1 km) the shadow and object size are the same. The outer solar system objects (TNOs and Centaurs) are the most distant minor bodies that we can observe, and they have -for sizes comparable to the closer main belt asteroids (MBAs)– significantly smaller angular sizes (see Table 2). However, the shadow sizes are independent of the object's distance from the earth, because the occulted stars are effectively at infinity (see Figure 1). The combination of uncertain ephemerides and small sizes for the Centaurs and TNOs makes them extremely challenging targets for predictable stellar occultations.

A third factor limiting our ability to predict stellar occultations is the astrometric uncertainty of the stellar catalogs. The most accurate catalogs currently available are UCAC4, with uncertainties between 15 and 100 milliarcseconds (mas), depending on star magnitudes (Zacharias et al. 2013) and URAT1 a new, but incomplete, stellar catalog with up to ~ 2-3 times better astrometric precision (Zacharias et al. 2015). The situation will improve significantly with the future GAIA catalog, which is expected to have an accuracy of ~ 5-14 microarcseconds ($\mu$as) for stars brighter than V = 12 mag, increasing to ~ 24 $\mu$as (for V= 15 mag) and ~290 $\mu$as (for V= 20 mag). This will greatly reduce the uncertainties in stellar occultation predictions in two ways: i) it will be possible to refine the orbit determinations using historical images of a particular TNO/Centaur and refine the astrometry using the GAIA catalog; ii) once target and star can be imaged together in the same field of view –FOV– of the image (a few weeks prior the occultation) it will also be possible to refine the prediction itself. Of course, the most accurate occultation predictions will make use of both of these techniques.

For occultations from JWST, an additional source of uncertainty in occultation predictions is the position of the telescope in its orbit near L2, 1.5 million kilometers from Earth. The expected accuracy of the JWST ephemeris is currently under study, and will be a key factor in estimating how well one can predict stellar occultations well in advance of the event. For minor bodies, we estimate that we require predictions accurate to ~300 km a full year in advance of the event, to allow for detailed proposal planning, and to better than 100 km a month in advance of the event to refine the details of the observation sequence. It is expected that JWST will perform station-keeping maneuvers roughly every 21 days, and the current estimated positional uncertainty is about 100 km one month in advance (see Section 3). Updates to the observation design can be accommodated easily up to about one week prior to the event (on-board schedules will be updated weekly during routine operations), and could be accommodated with as little notice as 24 hours (the minimum warning allowed for triggering a ToO observation). By comparison, Herschel Space Observatory was always within 10 km of the predicted orbit over a single day (24h) and the deviation was up to 300 km over a month. Herschel had correction maneuvers every 2 months, with delta-V typically ~0.1-1 $m \cdot s^{-1}$. A successful observation of a stellar occultation requires that the JWST location be known to better than the uncertainty on the path of the shadow in the sky plane, and to less than the shadow radius. With the currently expected JWST orbit accuracy the uncertainties in the stellar occultation predictions will be dominated by uncertainties in the knowledge of the JWST orbit for the smaller and more distant targets (see Table 2).

Table 2.-Apparent sizes and target/shadow diameters for different solar system bodies as observed from JWST.

| Body | R_JWST (AU) | Angular Size | Target Diameter (1000 km) | Parallax Uncertainty (mas) |
|---|---|---|---|---|
| **Jupiter** | 4.2 | 45.9” | 139.8 | 32.8 |
| **Saturn** | 8.5 | 18.9” | 116.5 | 16.2 |
| **Uranus** | 18.2 | 3.8” | 50.7 | 7.6 |
| **Neptune** | 29.0 | 2.3” | 49.2 | 4.8 |
| **Pluto** | 38.4 | 82.8 mas | 2.31 | 3.6 |
| **TNO** | 39.0 | 14.1 mas | ~0.4 | 3.5 |
| **Centaur** | 16.6 | 8.3 mas | ~0.1 | 8.3 |

**R_JWST (AU)** is the mean distance of the body to the JWST expressed in astronomical units. **Angular Size** is the mean apparent (angular) size of the body. **Target Diameter (1000 km)** is the mean diameter of the target, expressed in kilometers x 1000, which, in a stellar occultation, is equal to the projected shadow diameter. **Parallax Uncertainty (mas)** is the parallax error or the pointing uncertainty, expressed in milliarcseconds, due to the currently expected uncertainty of the JWST orbit over a month (~ 100 km).

Setting aside for the moment the challenges of accurate predictions for stellar occultations, this observing technique has the prospect of determining with high precision the sizes, shapes, and albedos of small bodies by measuring the times of disappearance/reappearance of a star behind the limb of the object. A single chord will provide a lower limit for the size of the object (in some cases, a single chord, together with an accurate astrometry, can even provide an equivalent diameter, e.g. Alvarez-Candal et al. 2014), whereas multiple chord events permit a fit for its shape and size. Unfortunately, for occultation observations of solar system objects from JWST, it is likely that only single chord occultations will be observed, since the typical event geometry will not provide the opportunity to observe the same occultation from earth-based observatories. Possible synergies with the Stratospheric Observatory for Infrared Astronomy (SOFIA) are currently under study, but it is not an easy task due to the difficulty of accurately predicting a stellar occultation from SOFIA, with flight schedules usually affected by meteorology, and to the very low probability that an occultation visible from SOFIA is also visible from JWST.

However, even a single chord occultation can be scientifically valuable. For example, a single chord observation could detect localized or global atmospheres at the nbar level (Sicardy et al. 2011; Ortiz et al. 2012). With sufficiently high SNR, it could allow the retrieval of vertical profiles of atmospheric pressure and temperature. Last but not least, central flashes due to the presence of an atmosphere can be detected if the single chord is a central (relative to the body) or close to central chord. The advantage of observing from space with JWST would make it easier to observe such a flash compared to a ground-based telescope. Single chord occultations also provide the opportunity for serendipitous discovery and characterization of satellites and rings around small bodies, such as the recently detected ring system orbiting the Centaur Chariklo (Braga-Ribas et al. 2014; Duffard et al. 2014) and probably the Centaur Chiron (Ortiz et al. 2015). For the airless (i.e. smaller) bodies the requirements on the knowledge of the JWST orbit are very tight (see Table 2 and Section 3). Nevertheless, each size measurement is still of keen interest, in particular for TNOs with sizes ≤ 100km which are unknown/unexplored and help to constrain the collisional history of the outer solar system. For the larger bodies, the requirements on knowledge of JWST orbit are looser because of the larger shadow size of the target. Some of these bodies (such as the dwarf planet Eris) are large enough to retain bound atmospheres (Schaller & Brown 2007), and the very high SNR provided by JWST could be uniquely able to detect them, or provide strong upper limits on any atmosphere for a non-detection, as explained above. In short, stellar occultations provide the opportunity to distinguish very fine structures or details not possible by direct imaging techniques, and in some cases rival or even exceed those obtainable from imaging observations from spacecraft that fly by or orbit the target bodies.

NIRCam is the JWST instrument best suited to observe stellar occultations. The achievable ~6.7 Hz cadence using a 64x64 subarray (2” or 4” FOV in the shortwave or longwave channel, respectively) provides nearly diffraction-limited km-scale resolution of minor body occultations. The wide coverage in wavelengths (from 0.7 to 4.8μm) is another advantage of using JWST-NIRCam to observe stellar occultations, because the optimum filter can be chosen to maximize the SNR of the occulted star and to minimize the light reflected by the object. In this sense, we consider four different end-member surfaces among the icy distant minor bodies (see Figure 2): i) objects with methane ice rich surfaces (*e.g.* Pluto, Eris, Makemake); ii) objects with water ice rich surfaces (*e.g.* Haumea Family); iii) objects with surfaces rich in organics (*e.g.* some red TNOs and Centaurs) and; iv) objects with surfaces dominated by silicates (*e.g. s*ome Centaurs and Trojans). Other TNO surfaces are likely to be combinations of these end-member surfaces. In general, for objects with methane ice dominated surfaces the F335M filter is the best choice, for objects with water ice dominated surfaces the preferred is the F300M (or any other with wavelength > 3 μm), for objects with organic rich surfaces F335M or F360M (or any other with wavelength > 3 μm). For objects with silicate dominated surfaces filters near or shortward of 1μm may be best, depending on stellar spectral type (see Table 3). To decide the optimum filter for any particular target one would need to take into account the reflectance spectrum of the object (e.g. water + CH4, CH4 + organics) and the spectral type (color) of the star, choosing the filter that maximizes the SNR on the star.

Table 3.- Best-choice filter(s) for TNOs and Centaurs.

| Main Component | Representative Objects | Best-choice Filter |
|---|---|---|
| $CH_4$ ice | Pluto, Eris, Makemake | F335M |
| $H_2O$ ice | Haumea & family | F300M (others with $\lambda > 3$ $\mu$m) |
| Organics | Some red TNOs/Centaurs | F335M, F360M (others with $\lambda >$ 3 $\mu$m) |
| Silicates | Some Centaurs/Trojans | $\lambda$'s ~ 1 $\mu$m or shorter |

Best-choice filter(s) for different end-member surfaces among the TNOs and Centaurs. The preferred filter is chosen taking into account the dominant composition spectrum and the transmission curves of the NIRCAM filters (see Figure 2). Main Component: main surface component.

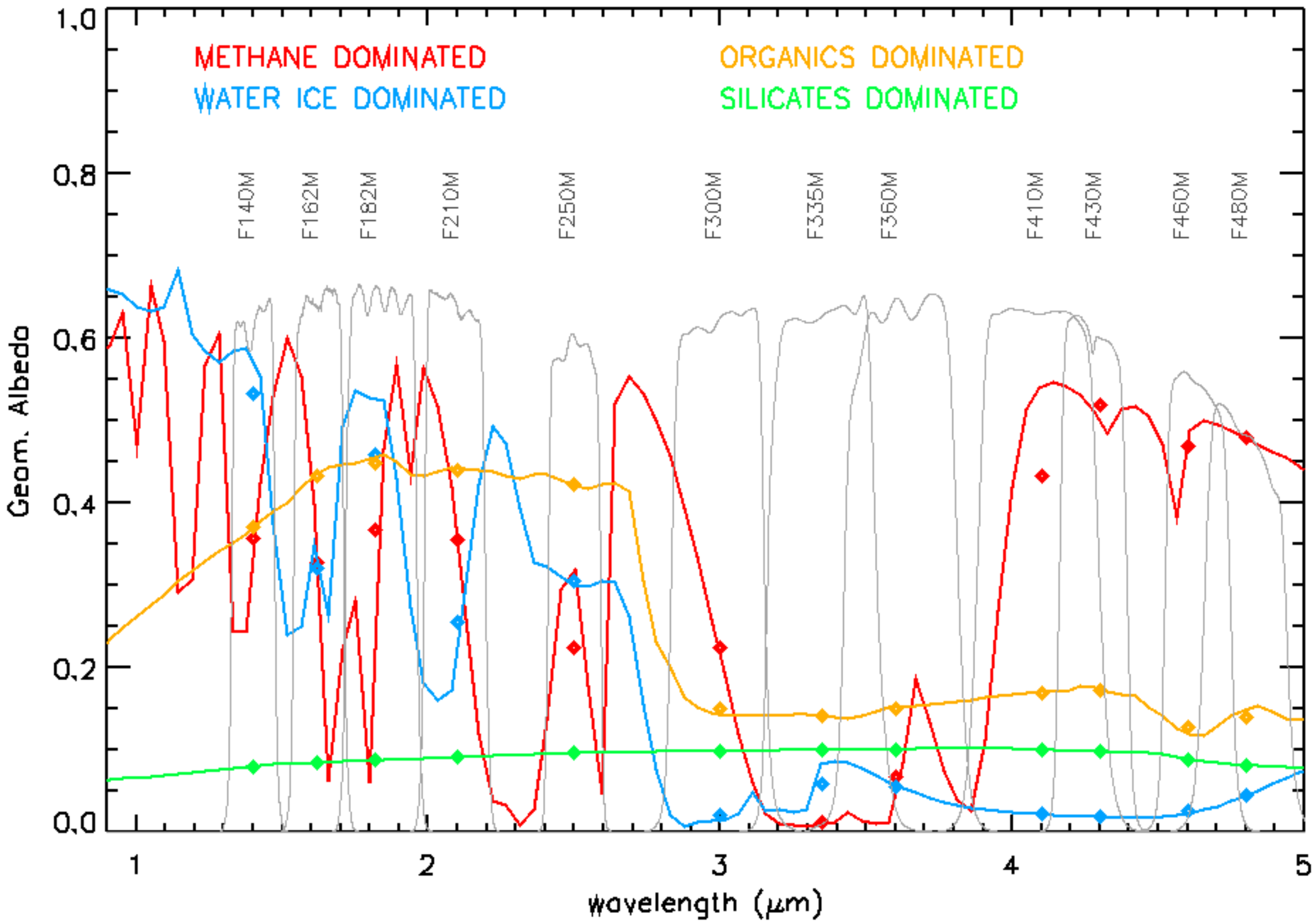

Figure 2.-Geometric Albedo of icy surfaces in the NIRCAM wavelength range. Colored lines show the geometric albedo of icy surfaces typical among the TNOs and Centaurs. These albedos have been obtained using the Hapke theory (Hapke 1993) for the reflectance of light on multicomponent surfaces. A methane ice dominated surface could represent the albedo of some Icy Dwarf Planets dominated by volatiles (e.g. Licandro et al. 2006a, 2006b, Brown et al. 2005). Water ice dominated surfaces correspond to Haumea and the other objects in the family (Pinilla-Alonso et al. 2007, 2008, 2009; Brown et al. 2007). Surfaces dominated by organics could be associated with objects that are red at visual wavelengths, e.g. Pholus and Sedna (Cruikshank et al. 1998; Emery et al. 2007). Surfaces dominated by silicates are more typically found among some Centaurs and Trojans (Emery et al. 2011; Tegler et al. 2008). The colored points correspond to the convolution of each of the albedos with the transmission curves of the NIRCAM filters (represented in gray). This figure shows that the optimum filter to observe an occultation would be determined by the surface composition of the TNO or Centaur, and also dependent on the color of the star.

To examine the potential utility of JWST to study stellar occultations by distant solar system minor bodies, we searched for candidate events for 13 large and/or interesting TNOs and Centaurs. Table 4 summarizes the results of that initial assessment of appulses possibly visible from JWST. For this exercise we used the UCAC4 and URAT1 star catalogs, the JWST planning ephemeris, and restricted the search to dates between Dec 1, 2018 and Jan 31, 2023. We included all events within the field of regard of JWST[2] and with a miss distance of 100 mas or less, to account for uncertainties in the catalog star positions and proper motions. As is shown in Table 4, these occultation events are going to be rare, that is not a specific weakness not to look into stellar occultations with JWST since events will be no more infrequent for JWST than they are for any ground-based or space-based observatory. In this sense, JWST is as good as any other established observatory, so its relative usefulness depends only on its capabilities (sensitivity, time resolution, photometric stability, …), which will be exceptionally good for stellar occultation studies. Anyway, Ixion seems to be a particularly attractive target for occultations, since it will be passing through a dense star field in the Milky Way over the next decade or so, enhancing the chances for a successful occultation. It is relevant to remark that the GAIA catalog will significantly expand the list of candidate stars. Those stars will be fainter than those presented in Table 4 and so will yield lower SNR, but may significantly increase the number of potential occultation events. As the sensitivity of JWST is higher (better PSF, lower background, ...) than comparable earth-based telescopes, then the new GAIA catalog will be a win that is unique to JWST and even larger ground-based telescopes.

[2]The field of regard of JWST is the range of allowed angular separation between the target object and the sun, which must lie between 85 and 135 degrees.

Table 4.- Predicted occultation appulses for 13 outer solar system minor bodies.

| Body | K < 9 | 9 < K < 10 | 10 < K < 11 | 11 < K < 12 | 12 < K < 13 | 13 < K < 14 | 14 < K <16 |
|---|---|---|---|---|---|---|---|
| 136199 Eris | ___ | ___ | ___ | ___ | ___ | ___ | ___ |
| 136472 Makemake | 1 | ___ | ___ | ___ | 1 | ___ | 3 |
| 50000 Quaoar | ___ | 1 | 2 | 3 | 11 | 7 | ___ |
| 90377 Sedna | ___ | ___ | ___ | ___ | 1 | ___ | 2 |
| 84922 (2003 VS2) | ___ | 1 | 1 | ___ | 3 | 7 | 24 |
| 136108 Haumea | ___ | ___ | ___ | ___ | ___ | 1 | 3 |
| 225088 (2007 OR10) | ___ | ___ | ___ | ___ | ___ | ___ | ___ |
| 120347 Salacia | ___ | ___ | ___ | ___ | ___ | ___ | 4 |
| 90482 Orcus | ___ | ___ | ___ | ___ | ___ | 1 | ___ |
| 28978 Ixion | 4 | 4 | 8 | 16 | 21 | 5 | 1 |
| 10199 Chariklo | ___ | 1 | ___ | 3 | 1 | 4 | 2 |
| 60558 Echeclus | ___ | ___ | 1 | 4 | 5 | 13 | 41 |
| 2060 Chiron | ___ | ___ | ___ | 1 | ___ | ___ | ___ |

Predicted occultation appulses for 13 outer solar system minor bodies (TNOs and Centaurs), based on the UCAC4 and (incomplete) URAT1 catalogs and the nominal JWST ephemeris, for the period Dec 1, 2018 to Jan 31, 2023. All events within the JWST field of regard and with a miss distance of 100 mas or less are included. The number of events for each object is listed by K magnitude range of the occultation candidate stars. Ixion is passing across a dense star field in the Milky Way, accounting for the large number of possible events for this tiny object. It is important to note that these appulse predictions are just notional, because we cannot yet predict where JWST will be in its orbit accurately.

## 2.2.- **Occultations by rings**

The distinctive ring systems of each of the solar system's giant planets exemplify a broad diversity of form and structure, and individually and collectively they provide insight into their dynamical environments, their interactions with satellites, their cosmogonic origins and space weathering effects, the influence of non-gravitational forces, and even as living records of recent impact events. Although the rings of Jupiter, Saturn, Uranus, and Neptune have all been observed close up by spacecraft, there is still a great deal to be learned by continued detailed remote observations, even from the distance of the earth. In this section, we identify the key scientific opportunities for discovery provided by JWST occultation observations of giant planet and minor body ring systems.

Much of what we know about the very detailed structure of the rings of Saturn, Uranus, and Neptune has been revealed by stellar and radio occultations. Indeed, the ring systems around Uranus and Neptune were first discovered by occultations, and they still provide the highest spatial resolution of the rings, revealing very fine structures not visible in direct imaging. Saturn's broad and complex ring system has been explored quite extensively, beginning with the *Voyager* 1 and 2 encounters, continuing with high-SNR earth-based occultations, two HST occultation events, and more than a decade of stellar and radio occultations from *Cassini*, scheduled to continue until mid-2017.

With all of these observations in hand, what could JWST stellar occultation observations add to our scientific understanding of planetary rings? One important benefit would be to extend the time baseline of observations of non-circular and time-variable structures. These include precessing eccentric and inclined ringlets and gap edges, evolving ring structure and density waves associated with co-orbital satellites Janus and Epimetheus, and identification of free and forced normal modes, such as those present in the outer edge of Saturn's B ring, the many eccentric ringlets in Saturn's C ring and Cassini Division, and Uranus's γ and δ rings. Although the rings of Saturn and Uranus would provide the easiest occultation targets for JWST, the stability and persistence of Neptune's ring arcs is an active area of interest, and a combined approach of JWST imaging and occultation studies could be quite productive. To set the general scene, we show representative stellar occultation observations by the rings of Saturn (Figure 3-A) and Uranus (Figure 3-B).

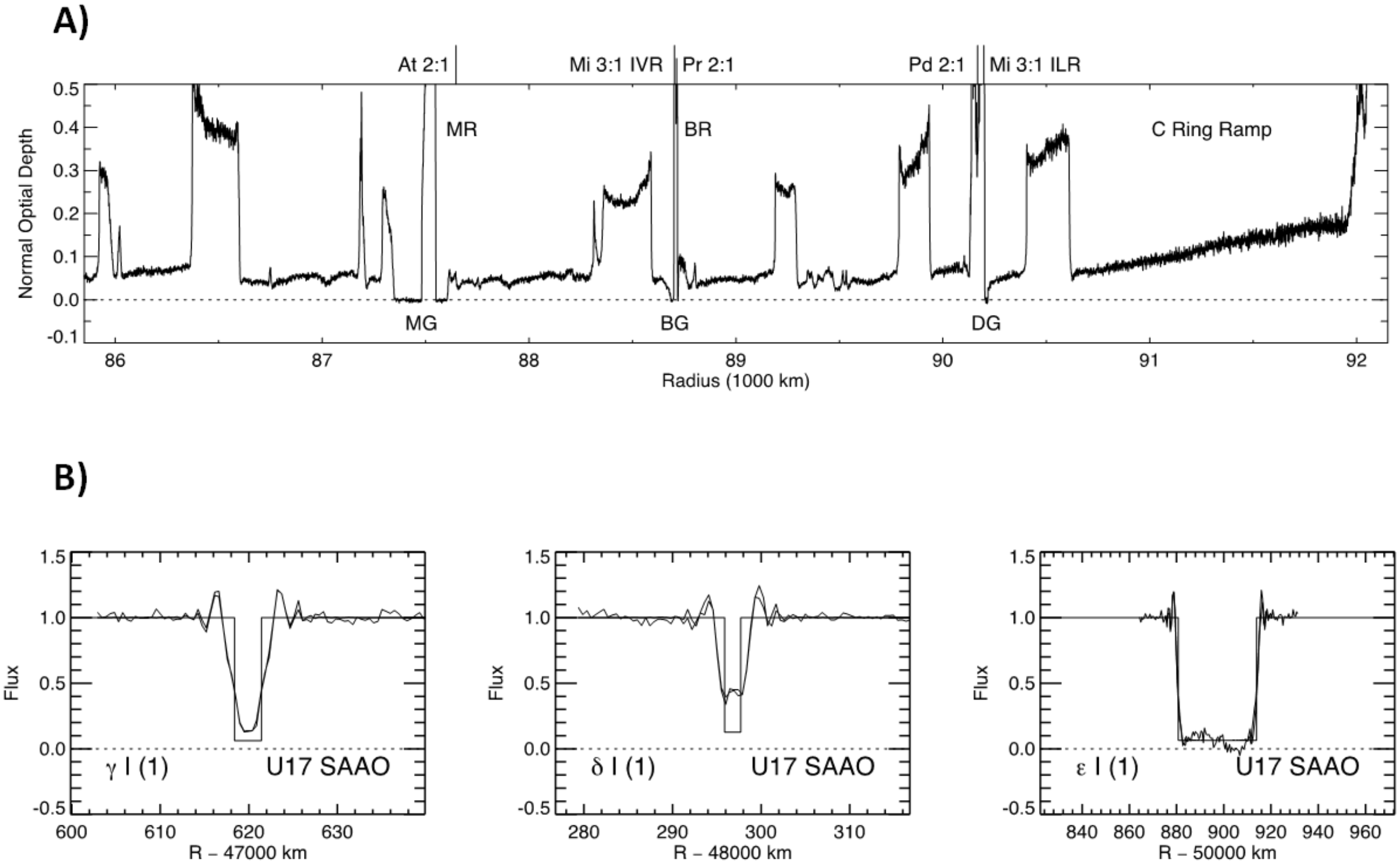

Figure 3.- **A)** Stellar occultation of Saturn's C ring from *Cassini* VIMS, showing detailed structure in the vicinity of three prominent ringlets: Maxwell (MR), Bond (BR), and Dawes, and their associated gaps in the rings (MG, BG, and DG). (After Fig. 1 of Nicholson et al. 2014.) **B)** High-SNR earth-based occultation observations of three Uranian rings, from the South African Astronomical Observatory (SAAO) -see Elliot et al. (1987) for additional details. The gamma ($\gamma$) and delta ($\delta$) ring profiles are diffraction-limited, but model fits provide very accurate

estimates of the rings' widths, mean optical depths, and orbital radii. The highly eccentric epsilon (ε) ring is sufficiently broad to reveal non-uniform internal structure.

JWST has the capability to provide exceptionally high-quality occultation profiles of ring systems. Its large aperture and the space environment provide great sensitivity and very high SNR. High quality earth-based occultations require very bright stars, and such events are quite rare. JWST's ability to obtain high-SNR observations from fainter stars both greatly increases the frequency of useful events, and also reduces the smoothing effects resulting from the finite angular diameter of the occulted star. Even a few JWST ring occultation observations could yield a substantial scientific return. With appropriate IR filters, scattered light from the planet and rings can be minimized.

As was pointed out in Section 2.1, NIRCAM is also the preferred instrument to observe stellar occultations by rings, because it can provide kilometer-scale resolution of giant planet rings. Historically, many earth-based occultations have used the IR K-band (near 2.2 μm), home to a strong $CH_4$ absorption band that renders the giant planets quite dark in reflected sunlight. There is no ideal NIRCam filter that can duplicate this effect near 2.2 μm, but the NIRCAM F300M filter, with a central wavelength near 3 μm, is in the water ice band of Saturn's rings, and is near to a $CH_4$ absorption band as well. Other attractive choices are the F335M and F360M filters where Saturn itself is very dark. In general, every filter above F300M can be a good choice (Figure 4). The final choice will require a more detailed assessment of scattered light from the planet and rings, taking into account the actual event geometry and the sky-plane path of the star relative to the planet.

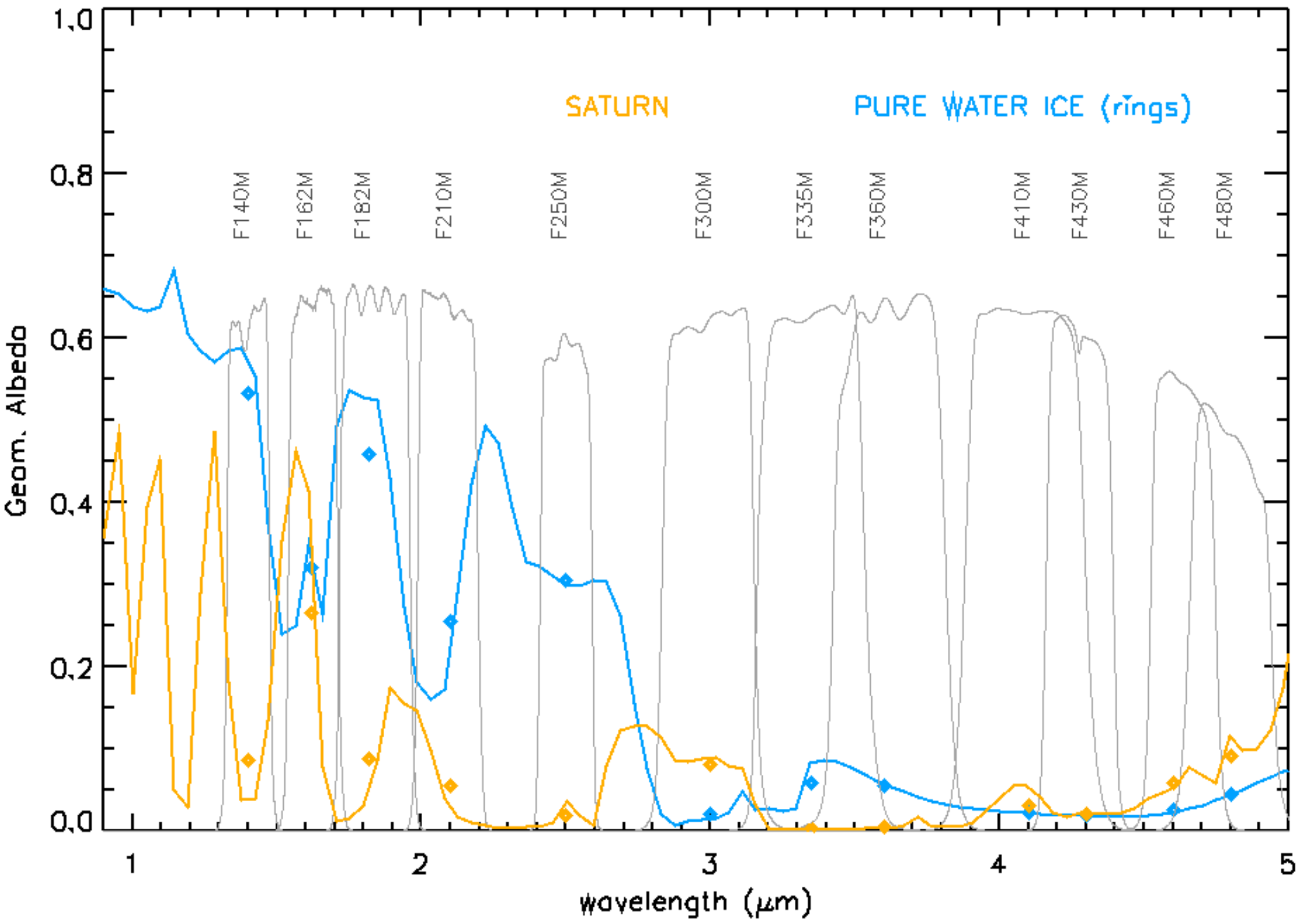

Figure 4.- The reflectance spectrum of Saturn* and water ice (pure model of water ice at 40 K), and the bandpasses of NIRCAM's medium filters. Note that the F300M appears to be well matched to the strong 3 μm ice absorption band in Saturn's rings, and Saturn is very dark in the F335M and F360M filters. In general, every filter above F300M is a good choice, although F300M, F460 and F480 have a slightly larger amount of light from the planet, background. *Albedo of Saturn estimated from combined data sets: Clark and McCord (1979): vis-nIR albedo spectrum; Karkoschka (1994): vis–1μm geometric albedo spectrum; Encrenaz et al. (1999): 2.5 – 5 μm ISO spectrum (Jy).

For completeness, we show Saturn's and Titan's IR absorption spectra in Figure 5. The F300M filter sensitivity is between 2.8 and 3.2 μm, where Saturn is not especially dark, but the disk of Saturn should be up to several arcsec away from the rings during much of a ring occultation, and if NIRCam's PSF is diffraction-limited, scattered light from the planet may be only a minor effect during ring occultations, although this will need to be studied in more detail.

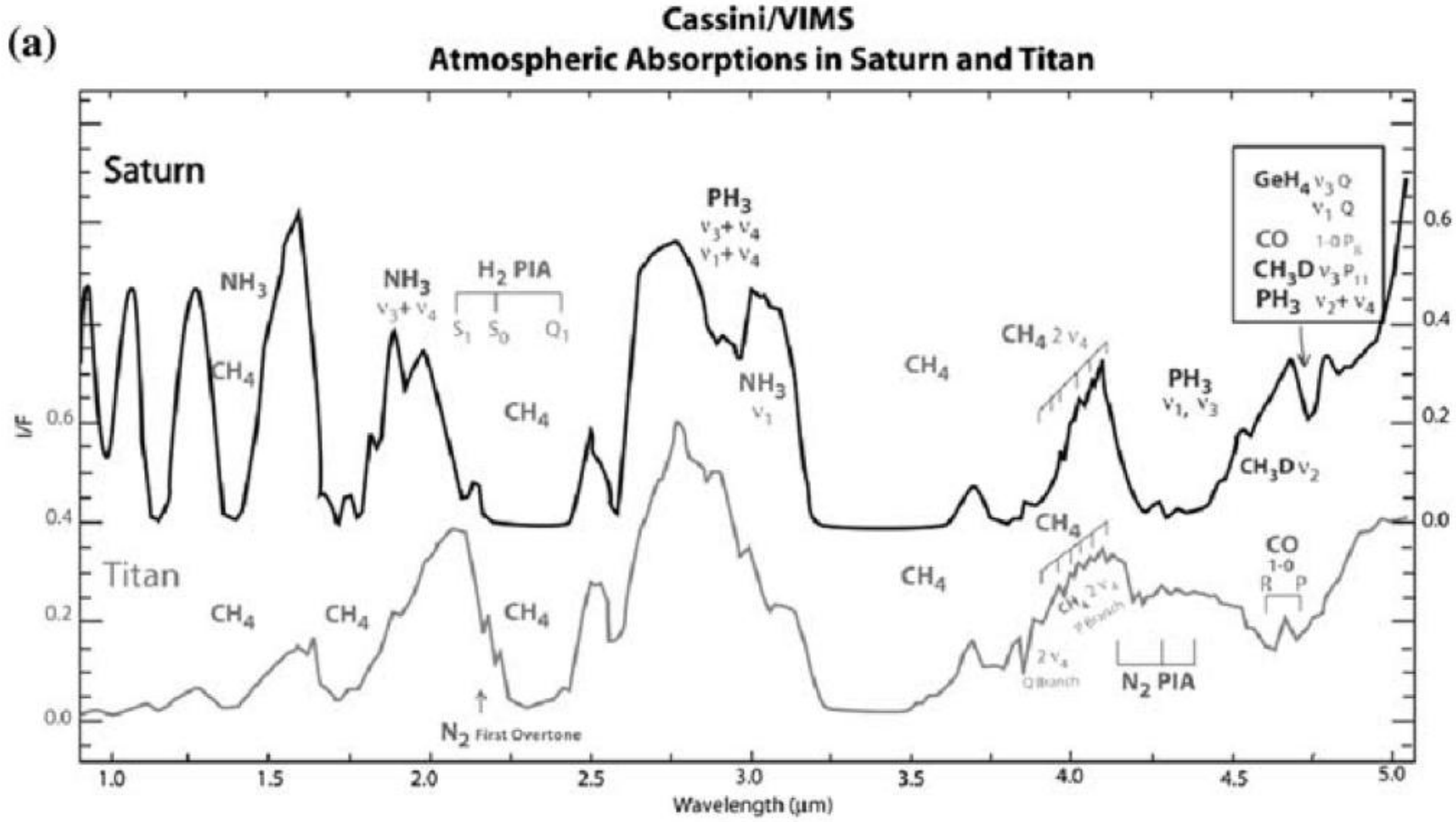

Figure 5.- Saturn's and Titan's atmospheric absorption spectra, from *Cassini* VIMS (from Baines et al. 2005). Note the strong $CH_4$ absorption band above 3.2 μm, which lies just above the NIRCAM F300M filter bandpass.

For giant planet ring occultations, the projected dimensions of the rings are of order 10,000–70,000 km. Even with uncertain JWST orbit determination and imprecise star positions, nominal event predictions well in advance can be useful for planning JWST and coordinated ground-based observing campaigns, for which a year of lead time is sometimes required.

As noted previously, the recent discovery of a ring system around Centaur Chariklo raises the possibility of using JWST to search for similar rings around other solar system minor bodies. For JWST to observe minor body ring occultations, the prediction challenges are quite extreme: the 100-1000 km scale ring systems have prediction uncertainties larger than the rings themselves (e.g. the ring system detected around Chariklo has orbital radii of 391 and 405 kilometers with widths of ~7 and ~3 kilometers respectively). Observation strategies will require a case-by-case assessment of the potential science return vs. the risk of a missed occultation.

The potential value of JWST ring occultations depends critically on their frequency. We have conducted a preliminary survey of giant planet ring occultation opportunities observable from the JWST. Using the current planning ephemeris for JWST and the 2MASS star catalog, we searched for stellar occultations that spanned the full radial range of the ring systems of Saturn, Uranus, and Neptune within the JWST's field of regard between Dec 1, 2018 and Jan 31, 2023 (the interval spanned by the available JWST planning ephemeris). Table 5 summarizes the number of predicted occultations, by K star magnitude range.

Table 5. Stellar occultations by planetary ring systems observable from JWST.

| Body | K < 9 | 9 < K < 10 | 10 < K < 11 | 11 < K < 12 | 12 < K < 13 | 13 < K < 14 | 14 < K < 15 |
|---|---|---|---|---|---|---|---|
| Saturn | 0 | 2 | 11 | 4 | 15 | 42 | 62 |
| Uranus | 0 | 0 | 0 | 0 | 2 | 3 | 3 |
| Neptune | 0 | 0 | 0 | 0 | 0 | 1 | 2 |

Number of stellar occultations by planetary ring systems observable from JWST between Dec. 1, 2018 and Jan. 31, 2023, within the listed K magnitude ranges, based on the 2MASS star catalog.

Below, we illustrate the geometry of three ring occultations predicted to occur in 2019. Figure 6-A shows the occultation geometry for a particularly attractive Saturn ring occultation of a K=11.71 magnitude star on April 5, 2019. The entire ring system is sampled during the occultation both on ingress and egress; tick marks are 30 minutes apart, and the duration of the occultation shown is 7 hours.

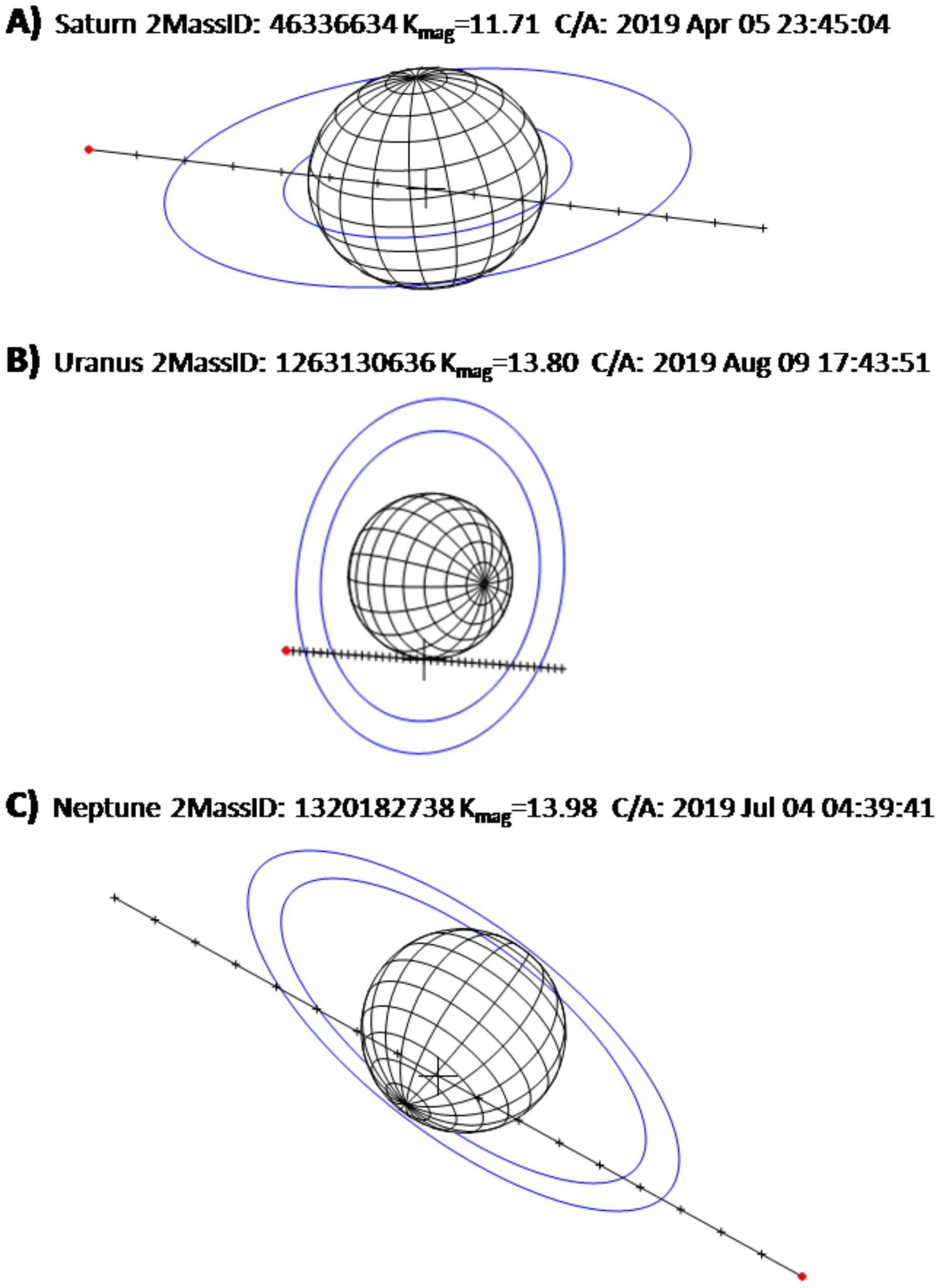

Figure 6.- **A)** Sky plane view of Saturn ring occultation viewed from JWST on April 5, 2019. The occulted star has K magnitude 11.71. Tick marks are 30 minutes apart. Ring system ingress begins at the red dot. **B)** Predicted sky plane track for a stellar occultation by the Uranus ring system as observed from JWST. The event occurs on Aug 9, 2019, and lasts for 20 hours (tick marks are 30 minutes apart). The resulting SNR of the observations would be quite high for a star of this magnitude (K= 13.8 mag), owing to the slow event velocity. Ring system ingress starts at the red dot. **C)** Occultation geometry of a predicted stellar occultation by Neptune's rings, as observed from JWST. The occulted star has K magnitude = 13.98, and the event occurs on July 4, 2019. Tick marks are 30 minutes apart, and ingress is marked by the red dot.

Figure 6-B shows the occultation geometry for one of the seven predicted Uranus occultations during this period. This is an unusually slow event, because Uranus is near a stationary point in its retrograde loop on the sky. The occultation lasts 20 hours, with a sky plane velocity of only 1.18 km/sec, resulting in exceptionally high SNR for a star of this brightness (K= 13.8 mag).

Finally, Figure 6-C shows the sky plane view of a predicted occultation by the Neptune ring system. Neptune events are relatively rare, owing to the paucity of bright stars in Neptune's vicinity and the relatively small subtended angle of the ring system. Adding to the challenge, Neptune's rings are incomplete, and a detectable drop in signal is likely to occur only if one of the ring arcs happens to be in the right position to occult the star during the event. (We have not included this constraint when calculating the occultation statistics for Neptune in Table 5).

Based on these preliminary predictions, we conclude that appealing stellar occultations by giant planets ring systems visible from JWST are abundant for Saturn, roughly annual for Uranus, and rather rare for Neptune. An important next step is to perform proper SNR estimates using the actual filter band passes, stellar and planetary spectra, taking into account the expected scattered light from the planet and rings. As a point of comparison, past occultations by the Uranian rings with $K < 8$ mag have yielded excellent SNR with a 4-m telescope, limited by atmospheric scintillation and spatial smoothing due to IR chopping and the finite angular diameter of the occulted star. JWST would avoid both of these sources of noise, and of course have a much larger collecting area.

Looking further into the future, it is useful to take note of the changing opening angle of the giant planet ring systems as observed from JWST, shown in Figure 7. The Uranus system is becoming more and more open as seen from the earth, while Saturn is moving from northern summer solstice to ring plane crossing near 2025. Neptune's ring orientation is changing quite slowly, owing to the planet's long orbital period, and Jupiter's rings are nearly edge on throughout this period, a result of Jupiter's very small obliquity. Given the low optical depth of Jupiter's tenuous rings and the unfavorable viewing geometry, we have not identified any potentially useful JWST ring occultations by the Jovian rings.

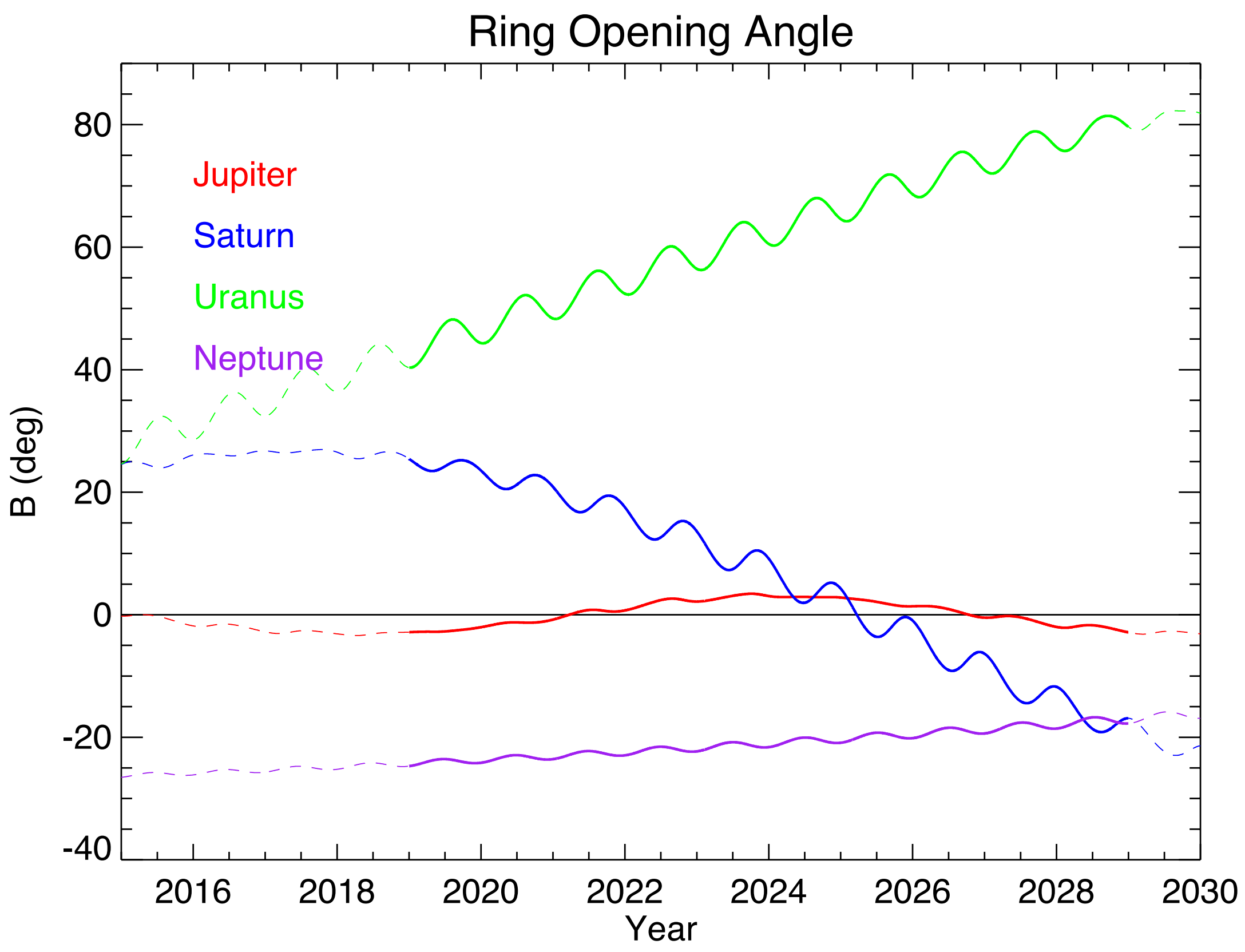

Figure 7.- Ring opening angle as viewed from the earth for the giant planet ring systems, plotted as a function of time. The solid curves span the decade 2019 to 2029, the expected JWST operational lifetime. The modulations in each curve result from the annual motion of the earth.

In addition to rings around giant planets, discovery observations of rings around minor bodies would have high scientific value. As noted previously (Section 2.1), the major challenges here are the prediction uncertainties for occultations by the bodies themselves, and the additional uncertainty of whether a ring is actually present and crosses the occultation path. We discuss the prediction uncertainties below, in Section 3.

## 2.3.- **Serendipitous stellar occultations**

The Fine Guidance Sensor (FGS) is a key component of the JWST attitude control system, and could be used for high-speed photometry as well. When science observations are in progress FGS will image a guide star every 80 msec (cadence of ~12 Hz) in an 8x8 subarray. Centroids from these images are used in real time to maintain pointing, and the centroid data produced on-board, including the count rate for the guide star images, will be sent to the ground. It is also possible to send the 8x8 subarray images themselves to the ground, but it is uncertain at this time whether downlink of the images will be done routinely.

We have used the JWST science operations design reference mission (SODRM) to analyze the properties of guide stars for slightly more than 1 year of science operations. The SODRM contains observations spanning the range of expected JWST science and observatory capabilities and is used for ground-system testing (including the guide-star selection system). The SODRM observations utilize over 17000 guide-stars. We find that the median photometric precision of the FGS photometry for the SODRM stars is ≤ 0.4%, and that over 70% of observing time will use guide stars with photometric precision ≤ 0.5% The remarkable combination of high cadence and high-SNR is enabled in part by the 0.6 –5 µm band-pass of FGS. Without the earth's atmospheric scintillation, the photometric precision is also expected to be excellent. We propose that one of the science projects for FGS should be to search for small size TNOs (< 1km) beyond Neptune and even further away by the observation of serendipitous stellar occultations of guide-stars produced by these objects (Roques et al 1987; Bailey 1976).

The direct detection of the smallest TNOs is a difficult task because they are extremely faint and are invisible to surveys using even the current largest ground-based and space telescopes. A powerful technique is to look for serendipitous transits of TNOs across distant stars. An occultation by a 1-km sized TNO located at 43 AU would produce an event duration ~ 0.3 seconds with a flux drop of ~ 10 %.Note that the flux does not drop to zero because the stellar occultation pattern by small objects at the outer solar system is diffraction-dominated (Roques et al. 1987). Such events could be readily detected in FGS data, as described above, and would offer important and unique constraints on the population of small TNOs.

As yet, there are few observational constraints on the size-frequency distribution of TNOs smaller than about 30 km because they are so difficult to detect in reflected light. Yet population statistics for small TNOs would allow us to investigate in more detail the collisional evolution history of the solar system. One way to study the collisional evolution is to examine the cumulative size distribution. The size distribution of TNOs has been accurately measured down to diameters of D > 30km (Fuentes et al. 2009; Fuentes & Holman 2008; Fraser & Kavelaars 2008; Fraser et al. 2008; Bernstein et al. 2004; Luu & Jewitt 2002). Although in reality, what has usually been measured is the $H_{mag}$ distribution (not the diameters distribution) and then, using assumed albedos, a size distribution has been derived. Attemps to actually measure size distributions have been made within the "TNOs are Cool" Herschel Space Observatory key program, as reported in Mommert et al. (2012) and Vilenius et al. (2012, 2014). The size distribution for large TNOs follows approximately a power law of the form:

$$dn/dD \propto D^{-q},$$

where the slope $q$ = 4.5 down to a diameter of $D$ ~ 90 km. A clear break in the size distribution has been detected near this 90 km size value (Fuentes et al. 2009; Fraser & Kavelaars 2009;

Bernstein et al. 2004), giving way to a shallower distribution for the smaller objects. Various models of the formation of the trans-neptunian belt that account for this break in the size distribution have been developed (Schlichting et al. 2013; Benavidez & Campo Bagatin 2009; Kenyon & Bromley 2001, 2004, 2009; Pan & Sari 2005; Benz & Asphaug 1999; Kenyon & Luu 1999a, 1999b; Davis & Farinella 1997; Stern 1996; Duncan et al. 1995), but these models make vastly different predictions of the size distribution for objects smaller than the break diameter. Thus, measurements of the size distribution of smaller objects (down to D ≥ 0.1km) that can be detected with FGS will be very useful to constrain and improve these size distribution models and to learn about the collisional history of the trans-neptunian belt.

A very similar approach to the one proposed here was employed by Schlichting et al. (2009) and (2012) using FGS on Hubble Space Telescope (HST). These authors detect 2 occultation events by objects ~ 500 m radius at 40-45 AU in ~ 31,500 star hours. However, an independent data set that could be compared to their results is still crucial. Based on our analysis presented above, JWST FGS will accumulate about 5000 hours of very high SNR guide-star photometry per year of science operations, or about 25000 hours over the 5-year mission lifetime requirement.The resulting data will be significantly more sensitive to serendipitous TNO occultations than the HST data were: the HST data had typical photometric precision of about 5%, 10 times worse than what we expect from JWST FGS data. However, it is unlikely that JWST will ever acquire 30,000 hours of guider data at moderate ecliptic latitudes as HST did, and the JWST sampling frequency (12 Hz) is significantly lower than HST's 40 Hz. On balance, we might expect JWST guider data to record a small but non-negligible number of serendipitious TNO occultations, depending on the (still quite uncertain) numbers of such objects. Such detections offer the possibilitiy of significant advances in our understanding of the size-distribution of TNOs in the collision-dominated size range.

Figure 8 shows the simulated change in flux for the occultation light curves for an object located at 43 AU with sizes from 0.1 to 1.0 km. The change in flux in the light curves is dominated by the diffraction pattern because the size of the object is smaller than the Fresnel scale (~ 4 km). A benefit from a random occultation observed in the NIR wavelength is that the duration of the physical event is larger compared with optical wavelengths and it is not sensitive to the irregular shape of the objects, but the disadvantage is that the drop in flux is smaller. Thus, high SNR guide stars (SNR ≥ 50) are preferred for searching serendipitous stellar occultation events by TNOs. The analysis presented at the beginning of this section suggests that JWST guide-star data will exceed this SNR goal significantly for nearly all guide stars.

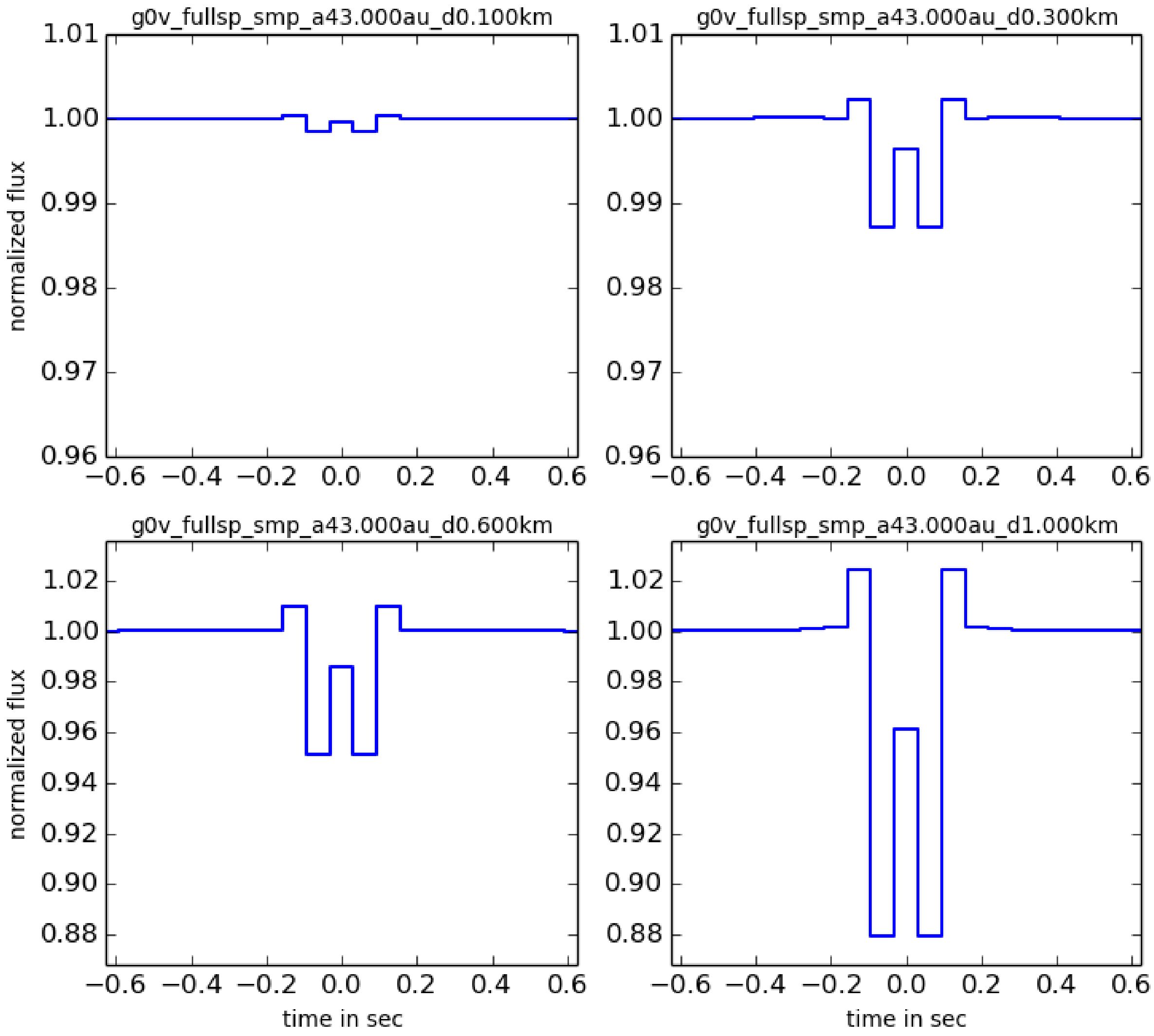

Figure 8.- Simulated light curves expected from a stellar occultation event produced by a TNO located at 43 AU for different body sizes, 0.1 km (top-left), 0.3 km (top-right), 0.6 km (bottom-left) and 1.0 km (bottom-right).

## 3. **JWST capabilities, justifications and needs**

As was stated in Section 2.1, knowledge of the predicted position of JWST as a function of time is critical for predicting occultation events for minor bodies (it is less critical for most giant-planet rings because of their large apparent size). The JWST flight-dynamics team continues to update their models for predicting the observatory ephemeris, having recently added components such as variable spacecraft attitude, uncertainty on station-keeping maneuvers, and statistically realistic pointing history models. These models currently suggest that the ephemeris can be predicted to an accuracy of 100 km out to 30 days from an orbit determination (a measurement of the position of the observatory), with the uncertainty in the position increasing linearly with time over that interval. This is accurate enough that it is reasonable to propose to observe minor body occultations with JWST, given the understanding that the observation could only be triggered once the predicted ephemeris uncertainty became smaller than the expected shadow size for the target and smaller than the uncertainty on the shadow-track itself. In order to propose occultation observations, the ephemeris uncertainty over a period of roughly one year is relevant (assuming a yearly proposal cycle for JWST). Over one year the current models suggest that the ephemeris uncertainty is 6000 km, with most of that error accumulating between about 50 and about 100 days. It is important to realize that such predictions will be statistical in nature even after the observatory is on station around L2. Because the sun shade acts as a solar sail, and its orientation with respect to the solar wind (and the strength of that wind) can't be known perfectly a priori, the significant non-gravitational forces acting on the observatory can't be accurately predicted. The event-driven operational model for JWST adds to the uncertainty because some observations that are on the schedule may be skipped, resulting in a deviation of the observatory attitude profile from what would be expected based on the uploaded observing plan. The JWST science and operations center will continue working with the flight-dynamics team to look for ways to improve the accuracy of observatory ephemeris predictions. For now it seems reasonable to expect that occultation observations for minor bodies, at least, will have to be submitted as target of opportunity (ToO) proposals.

Moving target tracking is not needed to observe stellar occultations from JWST, because we observe field ('static') stars that are occulted by solar system moving objects or rings. What is needed to perform stellar occultations is high time resolution (> 5 Hz) and sensitivity, which can be achieved with some of the JWST instruments. Taking this into account, our preferred instrument for predictable stellar occultations by solar system bodies and rings is NIRCam (see Sections 2.1 and 2.2). The variety of filters enables a choice that optimizes the SNR by reducing light scattered from the occulter itself, and/or from the host planet in the case of rings and satellites. The possibility to use the NIRSpec 1.6" square fixed slit to obtain occultation spectral is currently under study. On the other hand, FGS is the clearly preferred instrument for serendipitous occultations (see Section 2.3). We can obtain very valuable scientific results from this instrument as a by-product of other science projects.

Last, but not least, the release of GAIA catalogs will be very relevant for the better prediction of stellar occultation events, in particular for TNOs and Centaurs, and it will significantly expand, as well, the list of possible occulted stars to fainter stars only achievable, with enough SNR to get reliable photometry, with JWST (or with the largest ground-based telescopes). This is not as critical for giant planet ring events, or for serendipitous stellar occultations. A detailed explanation of how GAIA will improve the stellar occultation predictions is included in Section 2.1.

## 4.- Summary

- Stellar occultations by solar system (minor) bodies and rings are analyzed in terms of the JWST capabilities for three categories of events: i) predictable stellar occultations (we are mainly interested in occultations by distant minor bodies); ii) stellar occultations by rings; and iii) serendipitous stellar occultations by distant solar system minor bodies.

- The most challenging predictions for stellar occultations observable from JWST are the distant solar system minor bodies, such as TNOs and the Centaurs. The position of JWST in its L2 orbit must be known to an accuracy of around 300 km one year in advance to plan the proposals, and to better than 100 km one month in advance to refine the predictions. Without such JWST ephemeris accuracy, all the predictable stellar occultations by TNOs/Centaurs observed from JWST will be target of opportunity proposals.

- Light curves from stellar occultations observed from JWST will be single-chord, and will therefore provide only a lower limit for the object diameter. Possible synergies between JWST and other telescopes (like SOFIA), which allow obtaining multi-chord occultation, are currently under study. However, even a single-chord occultation can reveal very fine structures/details not possible by other techniques, such as: i) detection/characterization of atmospheres at the nbar level; ii) detection of central flashes due to an atmosphere, if the chord is central (or close to central) relative to the body; iii) serendipitous discovery/characterization of satellites and/or rings around small bodies.

- The future release of the GAIA stellar catalog will significantly improve the accuracy of stellar occultation predictions, particularly those involving distant solar system minor bodies. The accuracy of these catalogs will allow refinement of the orbits of particular TNOs/Centaurs, and will enable very accurate occultation predictions a few weeks prior the occultation. The GAIA catalog will significantly extend the list of possible occulted stars to fainter stars, which will be reachable with JWST.

- The preferred instrument to observe predictable stellar occultations by solar system bodies and rings is NIRCam due to its sensitivity and high time resolution. The best filters to observe stellar occultations should be chosen as a compromise between the larger SNR of the occulted star and the smaller light scattered from the occulted object or nearby planet (see Table 3, and Figures 3 and 6). The feasibility of using NIRSpec for occultation observations should also be explored.

- The preferred instrument to observe serendipitous stellar occultations is FGS due to its 12 Hz cadence and the large amount of star hours that can be obtained from its observations. It would allow extracting relevant information about the small population of TNOs/Centaurs from guide data for at least 70% of guide stars. We propose using the FGS data from other JWST projects to get this information as a very scientific valuable by-product.

- We identify appulses for stars with $K_{mag} < 16$ mag visible from JWST from Dec 1, 2018 to Jan 31, 2023 for 13 relevant TNOs and Centaurs. These appulses have been obtained using UCAC4 and URAT1 star catalogs.

- We identify stellar occultations by rings of Saturn, Uranus and Neptune visible from JWST from Dec. 1, 2018 to Jan. 31, 2023 (occultations by Jupiter's rings are not favorable during this period because they are nearly edge-on). Several of these events are analyzed in detail in this work. Particularly relevant are the predicted occultations by the incomplete Neptune rings. The possibility to discover rings around solar system minor bodies (as those discovered around centaurs Chariklo and Chiron) is also discussed.

- Serendipitous stellar occultations light curves, as observed by JWST-FGS, are simulated for different objects located at 43 AU from the sun, for sizes between 0.1 and 1 km. Our modelling indicates that FGS guide star data will allow high-SNR (≥ 200) detections of such occultations of the majority of guide stars.
- Last, but not least the predictable stellar occultations by solar system minor bodies require very little telescope time, while serendipitous stellar occultations can be obtained as a by-product of other projects: both kind of observations have a low observational cost and a very important scientific return.